\documentclass[11pt]{kimreview}

\usepackage{bm}
\usepackage{amsmath}
\usepackage{amssymb}
\usepackage{epstopdf} 
\usepackage{wrapfig}  
\usepackage{cite}

\begin{document}


\title{On the Atomic Cluster Expansion: interatomic potentials and beyond}


\author{Christoph Ortner}
\affiliation{Department of Mathematics, University of British Columbia, Canada}



\publishyear{2023}
\volumenumber{1}
\articlenumber{01}
\submitdate{July 18, 2023}
\publishdate{August 8, 2023}
\doiindex{10.25950/c7f24234}
\doilink{10.25950/c7f24234}


\paperReviewed
{Atomic cluster expansion for accurate and transferable interatomic potentials}
{R.\ Drautz}
{\href{https://doi.org/10.1103/PhysRevB.99.014104}{Phys.\ Rev.\ B, 99:014104 (2019)}}


\maketitle



\begin{abstract}
The Atomic Cluster Expansion (ACE) \cite{Drautz2019-er} provides a systematically improvable, universal descriptor for the environment of an atom that is invariant to permutation, translation and rotation. ACE is being used extensively in newly emerging interatomic potentials based on machine learning. This commentary discusses the ACE framework and its potential impact.
\end{abstract}
\medskip


\section*{Background}

Machine-learning interatomic potentials (MLIPs) are a novel class of potential energy models that are beginning to revolutionize atomistic-scale materials and molecular simulation. For an increasing number of systems of scientific and technological interest they are capable of closely matching the predictions of high fidelity electronic structure models, at a tiny fraction of the computational cost. There is legitimate optimism in the MLIPs community that this class of models will, in the coming years, enable routine large-scale and long-time simulations at similar accuracy as electronic structure models. I refer to \cite{DeringerCsanyi2021ChemRev,musil2021physics,behler_csanyi_2021} for recent overviews of the field.

This commentary focuses on the {\it Atomic Cluster Expansion} (ACE), a specific MLIP flavour introduced by Drautz~\cite{Drautz2019-er}. Despite some effort at objectivity, it emphasizes my personal perspective and taste.

MLIPs are conceptually not too different from the empirical interatomic potentials of the past several decades: one chooses a parameterized functional form for the total potential energy, incorporating as much physical knowledge as one can, then estimates the remaining parameters from available data. The step-change introduced by MLIPs is that they employ far more flexible parameterisations\footnote{Ideally, MLIPs should be systematically improvable, or universal in the language of machine learning; however, many MLIPs cannot be classified as such.} and are therefore in principle able to capture a much wider range of physics at higher accuracy. This was made possible by the rapid increase in computing resources and resulting readily available {\it ab initio} simulation data.

MLIPs are best thought of as a class of models that borrow machine learning ideas to fill the gaps left by empirical modelling and make them systematically improvable. Such work requires a fine balance between flexibility of the models and strong physical priors, and this is precisely where the ACE model and its relatives show their strength. Indeed, according to Drautz\footnote{personal communication}, his original motivation for developing the ACE model was to demonstrate that it is possible to construct interatomic potential models that are systematically improvable ({\it universal}, in the language of machine learning) while maintaining physical interpretability.

\section*{ACE as an interatomic potential model}
\def\Nmax{{\bar{N}}}
\def\N{N}
It is a decades-old procedure to expand total energy into site energies, $E = \sum_i E_i$, and then site energies into many-body terms,
\begin{equation}  \label{eq:naive_manybody}
\begin{aligned}
    E_i =\,& V_0(Z_i) +
    \sum_{j_1} V_1({\bm r}_{ij_1}, Z_{j_1}; Z_i)
    + \sum_{j_1 < j_2} V_2({\bm r}_{i j_1}, Z_{j_1}, {\bm r}_{i j_2}, Z_{j_2}; Z_i)
    + \dots   \\
    & + \sum_{j_1 < \dots < j_\Nmax}
    V_{\Nmax}\big( {\bm r}_{i j_1}, Z_{j_1}, \dots, {\bm r}_{i j_\Nmax}, Z_{j_\Nmax}; Z_i),
\end{aligned}
\end{equation}
where $({\bm r}_i, Z_i)$ are atom position, atomic number pairs, ${\bm r}_{ij} = {\bm r}_i - {\bm r}_j$ are relative atom positions and the summation over indices $j_t$ ranges only over atoms within some predefined cutoff radius.

Even if we could formulate many-body potential terms $V_N$ that can be evaluated at $O(1)$ cost, the combinatorial scaling of $\sum_{j_1 < \dots < j_\Nmax}$ makes this approach entirely impractical for $\Nmax > 3$. One of the key innovations of the ACE model~\cite{Drautz2019-er} was to rewrite the many-body expansion as
\begin{equation}  \label{eq:manybody}
\begin{aligned}
    E_i =\,& U_0(Z_i) +
    \sum_{j_1} U_1({\bm r}_{ij_1}, Z_{j_1}; Z_i)
    + \sum_{j_1, j_2} U_2({\bm r}_{i j_1}, Z_{j_1}, {\bm r}_{i j_2}, Z_{j_2}; Z_i)
    + \dots   \\
    & + \sum_{j_1, \dots, j_\Nmax}
    U_{\Nmax}\big( {\bm r}_{i j_1}, Z_{j_1}, \dots, {\bm r}_{i j_\Nmax}, Z_{j_\Nmax}; Z_i);
\end{aligned}
\end{equation}
that is, summation is not only over unique clusters as in \eqref{eq:naive_manybody} but over repeated $N$-clusters as well as ``spurious'' clusters with repeated atom indices (self-interaction). After expanding the potentials $U_\N$ in a tensor product polynomial basis, transformed to allow only rotational and reflection invariant basis functions, this self-interacting expansion leads to a remarkably simple and efficient four-stage evaluation scheme:
\begin{equation} \label{eq:ACEscheme}
    A_{nlm}^{(i)} = \sum_j R_{nl}(r_{ij}, Z_j, Z_i) Y_l^m(\hat{\bm r}_{ij}),
    \quad
    {\bf A}_{\bf nlm}^{(i)} = \prod_{t = 1}^{\N} A_{n_t l_t m_t}^{(i)},
    \quad
    {\bf B}^{(i)} = \mathcal{C} {\bf A}^{(i)},
    \quad
    E_i = {\bm \theta} \cdot {\bf B}^{(i)}.
\end{equation}
Details and an in-depth performance assessment can be found in \cite{Drautz2019-er,Lysogorskiy2021-wa,DUSSON2022}.

The evaluation cost scales only linearly in the number of neighbours, unlike the combinatorial scaling of a naive cluster expansion. It is systematically improvable ({\em universal}) in the limit of taking the cutoff radius, body-order and polynomial basis to infinity. This combination of performance, universality and interpretability of the model as a many-body expansion rapidly captured the attention of the wider MLIPs community; see, e.g., \cite{NICE,vandermause2022,hyperactive2022,Goff2022-wt,kovacs2021} for a few selected examples building directly on~\cite{Drautz2019-er}. Seko, Togo and Tanaka~\cite{2019_Seko} independently developed a method that shares many ideas with \cite{Drautz2019-er}.

In my personal view, the greatest strength of ACE is its flexibility. The basic variant described in \eqref{eq:ACEscheme} is a {\em linear model}, which makes it attractive for incorporating it into a Bayesian framework, e.g., for an active learning type workflow~\cite{vandermause2022, hyperactive2022}. However, one can also employ the many-body expansion in several imaginative ways, e.g., extending the Finnis-Sinclair model which results in a site energy of the form $E_i = E_i^{(1)} - \sqrt{E_i^{(2)}}$, where each $E_i^{(p)}$ is a scalar expanded as in \eqref{eq:ACEscheme}. After also parameterizing the radial basis $R_{nl}(r_{ij}, Z_j, Z_i)$ (making it trainable), we obtain a flavour of ACE made available in the {\tt PACEMAKER} open source software package~\cite{Bochkarev2022-ky}. Drautz also extended the method in a straightforward fashion to the parameterisation of equivariant material properties such as charge or magnetic dipoles~\cite{Drautz2020-mg} and to multi-layer models with learnable atomic environment features that are themselves parameterized by ACE~\cite{Bochkarev2022-ky}. I will give further examples below.

Although only very recently released, the {\tt PACEMAKER} package has already been used successfully to parameterize a number of state-of-the-art interatomic potentials for materials, including, but not limited to Cu~\cite{Lysogorskiy2021-wa}, C~\cite{Qamar2023-vh}, Mg~\cite{Ibrahim2023-ni}, Fe~\cite{Rinaldi2023-nb} (with extension to account for magnetism) and Pt-Rh nanoparticles~\cite{Liang2023-sz}.

\section*{Connections}
To conclude the discussion of {\it ACE as an interatomic potential}, I briefly put it into context of a few prior methods with which it shares similarities. Bowman and Braams' permutation-invariant polynomials~\cite{Braams2009-bj} formulate an expansion of the {\em total} potential energy in terms of isometry and permutation-invariant polynomials using computational invariant theory methods. ACE~\cite{Drautz2019-er} can be seen as an explicit and computationally more efficient way to construct an invariant expansion.
Bartok, Payne, Kondor and Csanyi~\cite{bartok2010gaussian} introduced a systematic approach to expand of the atomic environment (SOAP) and then used a kernel method to represent site energies $E_i$. The SNAP method of Thompson et al.~\cite{THOMPSON2015SNAP} extends the SOAP idea from 3-body descriptors (power spectrum) to 4-body descriptors (bi-spectrum). ACE~\cite{Drautz2019-er} can be understood as an extension of the SOAP and SNAP descriptors to arbitrary high correlation orders, resulting in a {\em complete linear basis} and thus allowing expansion of the site energy as a linear model \eqref{eq:ACEscheme}. Moment tensor potentials (MTPs)~\cite{Shapeev2016-pd} construct an alternative complete linear polynomial basis for site energies. The key difference is that MTP employs tensor contractions in Euclidean coordinates, while ACE employs tensor contractions in a spherical harmonics representation. This seemingly superficial but possibly crucial difference appears to drive the relatively fast adoption of ACE. Employing an irreducible representation of the rotation group makes it easier to select an appropriate set of descriptors. This also makes the ACE formulation attractive for the quantum modelling community used to working with spherical harmonics representations.

\section*{ACE as a general many-body method}
Aside from a method to develop MLIPs, ACE can also be understood as a general methodology to incorporate many-body interaction into physical models in an interpretable, systematic and computationally efficient way. With relatively straightforward adaptions it was successfully used for parameterizing tight binding hamiltonians~\cite{2021-acetb1,Nigam2022-wy}, wave functions via the backflow transformation~\cite{2022-wave1,Zhou2023-ro}, and jet-taggers~\cite{Munoz2022-ye}. Another potentially far-reaching application is the integration of many-body interaction into equivariant deep neural network architectures. First, the ACE formulation was employed in \cite{botnet} to provide new theoretical perspectives on the E3NN and NequIP~\cite{nequip} architectures. This analysis led to two new model architectures, ALLEGRO~\cite{Musaelian2023-sf} and MACE~\cite{MACE2022}, which at the time of writing this commentary represent the state-of-the-art in MLIPs accuracy.


\section*{Conclusion}
The Atomic Cluster Expansion (ACE)~\cite{Drautz2019-er} is a canonical formulation for many-body parameterisations of equivariant properties of particle systems. Due to its crystal clear interpretability, excellent performance, accuracy, and adaptability to new problems, it has established itself as one of the main pillars of scientific machine learning for molecular dynamics simulation, and is just beginning to also impact other fields. Its greatest success so far is in parameterising interatomic potentials for materials simulation.
The discovery of the ACE formalism is a major advance in the evolution of interatomic potentials for large-scale atomistic materials modelling and MLIPs based on ACE will, over time, likely supplant venerable empirical interatomic potential models such as embedded atom and bond order potentials.


\bigskip

\bibliographystyle{unsrt}
\bibliography{kim-review-commentary-2023-08-01}

\end{document}